# Electric Switching of the Charge-Density-Wave and Normal Metallic Phases in 1T-TaS$_2$ Thin-Film Devices

A. Geremew[1], S. Rumyantsev[1,2], F. Kargar[1], B. Debnath[3], A. Nosek[4], M. Bloodgood[5], M. Bockrath[4], T. Salguero[5], R. K. Lake[3], and A. A. Balandin[1,*]

[1]Nano-Device Laboratory (NDL) and Phonon Optimized Engineered Materials (POEM) Center, Department of Electrical and Computer Engineering, University of California, Riverside, California 92521 USA

[2]Center for Terahertz Research and Applications (CENTERA), Institute of High-Pressure Physics, Polish Academy of Sciences, Warsaw 01-142, Poland

[3]Laboratory for Terascale and Terahertz Electronics (LATTE), Department of Electrical and Computer Engineering, University of California, Riverside, California 92521, USA

[4]Department of Physics and Astronomy, University of California, Riverside, California 92521 USA

[5]Department of Chemistry, University of Georgia, Athens, Georgia 30602 USA

---

[*] Corresponding author (A.A.B.): balandin@ece.ucr.edu ; web-site: http://balandingroup.ucr.edu/

## Abstract

We report on switching among three charge-density-wave phases – commensurate, nearly commensurate, incommensurate – and the high-temperature normal metallic phase in thin-film 1T-TaS$_2$ devices induced by application of an in-plane electric field. The electric switching among all phases has been achieved over a wide temperature range, from 77 K to 400 K. The low-frequency electronic noise spectroscopy has been used as an effective tool for monitoring the transitions, particularly the switching from the incommensurate charge-density-wave phase to the normal metal phase. The noise spectral density exhibits sharp increases at the phase transition points, which correspond to the step-like changes in resistivity. Assignment of the phases is consistent with low-field resistivity measurements over the temperature range from 77 K to 600 K. Analysis of the experimental data and calculations of heat dissipation suggest that Joule heating plays a dominant role in the electric-field induced transitions in the tested 1T-TaS$_2$ devices on Si/SiO$_2$ substrates. The possibility of electrical switching among four different phases of 1T-TaS$_2$ is a promising step toward nanoscale device applications. The results also demonstrate the potential of noise spectroscopy for investigating and identifying phase transitions in materials.

Switching between various material phases at room temperature by the application of electric field has the potential of becoming a new device paradigm for future electronic and optoelectronic technologies[1–4]. Among the promising material candidates, which must exhibit phase changes characterized by abrupt resistivity changes and hysteresis, is the 1T polymorph of tantalum disulfide (TaS$_2$). The quasi-two-dimensional (2D) van der Waals layered crystalline 1T-TaS$_2$ exhibits charge-density-wave (CDW) effects, *i.e.* periodic modulation of the charge density and the underlying lattice resulting from the interplay between the electron-electron and electron-phonon interactions[5–13,14]. The CDW state becomes fully commensurate with the lattice below ~200 K[15–17]. The commensurate CDW (C-CDW) consists of a $\sqrt{13} \times \sqrt{13}$ reconstruction within the basal plane that forms a star-of-David pattern in which each star contains 13 Ta atoms. The Fermi surface, composed of 1 *d*-electron per star, is unstable, so that the lattice reconstruction is accompanied by a Mott-Hubbard transition that fully gaps the Fermi surface and increases the resistance[15,18–21]. As the temperature increases above 180 K, the C-CDW phase breaks up into a nearly commensurate CDW (NC-CDW) phase that consists of ordered C-CDW regions separated by domain walls[22]. This C-CDW to NC-CDW transition is revealed as an abrupt change in the resistance with a large hysteresis window in the resistance profile at ~200 K. As the temperature is increased to ~350 K, the NC-CDW phase melts into an incommensurate CDW (I-CDW) in which the CDW wave vector is no longer commensurate with the lattice. This transition is accompanied by a smaller hysteresis window in the resistivity[23,24]. Only at high temperatures of ~500 K – 600 K does the I-CDW phase melt into the metallic normal phase (NP) of 1T-TaS$_2$[5]. Until now, the transition from the I-CDW to the NP phases has been achieved only via heating of 1T-TaS$_2$ samples.

Apart from temperature, various CDW phase transitions in 1T-TaS$_2$ and some related materials, *e.g.* 1T-TaSe$_2$, can be induced by external perturbations including doping[17,17,25–31] photoexcitation[8,32], pressure[5], carrier injection[6,33], reduction of thickness[34–36], electric field[9,24,33], laser pulse[8,37], and gate voltages[33,38,39]. The existence of several CDW phases, in addition to electrically driven switching among them within a short time scale, makes 1T-TaS$_2$ a particularly promising candidate for electronic device applications [6,33,40] even though the exact nature of these phases continues to be a subject of debate [41–44]. Major impetus for this research came from the demonstration of a 1T-TaS$_2$-based voltage controlled oscillator operating at room temperature (RT); this device utilizes the NC-CDW – I-CDW phase transition, which occurs at 350 K in the "low-bias" regime[38]. Here we refer to the resistivity as "low-bias" when it is measured at very low electric bias, *i.e.* few mV, to guarantee the absence of self-heating and thus preclude transitions induced entirely by the temperature change[38,45]. Subsequent work has focused on the implementation of such CDW devices for information processing[46,47] and radiation-hard applications[48,49].

It has been suggested that electric field and current not only interact with the CDW but also result in Joule heating, which in turn causes the phase transitions in 1T-TaS$_2$[50]. The extent to which the electrical field by itself, versus self-heating associated with the current conduction, is responsible for inducing transitions between C-CDW and NC-CDW or NC-CDW and I-CDW is still under debate[38,50,51]. Moreover, although the electric field induced C-CDW to NC-CDW and NC-CDW to I-CDW phase transitions in 1T-TaS$_2$ have been reported by many groups[6,38,52] there have been no reports on the electric-field induced transition to the normal metallic phase. This situation can be attributed to the difficulty of reaching temperatures of $T = 550$ K – 600 K via self-heating due to the applied electric bias and the less pronounced change in the resistivity at the I-CDW – normal phase (NP) transition. The possibility of using the I-CDW – NP transition for devices is attractive, however, due to their anticipated radiation-hardness,[48,49] which stems from the inherent tolerance of metallic materials to radiation damage compared with semiconductors.

In this Letter, we report on inducing the I-CDW – NP transition by applying an in-plane electric field in 1T-TaS$_2$ devices on conventional Si/SiO$_2$ wafers, operating over a wide temperature range, including RT. The measurements of the low-frequency electronic noise, *i.e.* current fluctuations, have been used to unambiguously observe the phase transitions, particularly the switching from the I-CDW to NP. The low-frequency noise, typically with the spectral density $S(f) \sim 1/f^\gamma$ ($f$ is the frequency and parameter $\gamma \approx 1$), is found in almost all materials and devices[53,54]. Although practical applications can benefit from the reduction of the low-frequency noise, it also can be used as a characterization tool that reveals information about the physical processes in materials[44,45,55,56]. The electrical switching among three CDW phases and one normal phase of 1T-TaS$_2$ is demonstrated with the help of low-frequency-noise spectroscopy. The rest of the paper is organized as follows. First, the material preparation, device fabrication, and resistivity measurements are briefly outlined. Second, low-frequency noise measurements are described in the context of monitoring phase transitions and carrier transport in CDW materials. Third, simulation of heat dissipation is used to assess the relative importance of the self-heating and electric field for inducing CDW and metallic phase transitions.

Multi-millimeter-sized crystals of 1T-TaS$_2$ were synthesized by iodine-mediated chemical vapor transport (CVT) (*Methods*) and their quality was evaluated by X-ray diffraction, energy dispersive spectroscopy, and other material characterization techniques, as previously detailed[38,45]. Thin layers of crystalline 1T-TaS$_2$ were cleaved by mechanical exfoliation and transferred to SiO$_2$/Si substrates. We intentionally used 1T-TaS$_2$ layers with thicknesses greater than 9 nm to preserve the C-CDW – NC-CDW phase transition. It is known that 1T-TaS$_2$ quasi-2D films with thicknesses approaching a monolayer do not reveal the C-CDW – NC-CDW transition[23,38]. The transmission line measurement (TLM) pattern was used for device fabrication. The metal Ti/Au contacts were evaporated through a shadow mask to avoid chemical contamination and oxidation during the fabrication process. After the contact evaporation, the devices were transferred to another vacuum chamber for transport measurement. A large number of 1T-TaS$_2$ devices were fabricated and tested. An optical microscopy image of a representative device together with the schematic of its layered structure are shown in Figure 1 (a). All current-voltage (I-V) characteristics and resistivities were measured in the cryogenic probe station (Lakeshore TTPX) with a semiconductor analyzer (Agilent B1500). The low-frequency noise experiments were conducted in the two-terminal device configuration but with different channel lengths. The noise spectra were measured with a dynamic signal analyzer (Stanford Research 785). Additional device fabrication details are provided in the *METHODS* and *Supplementary Materials*.

[Figure 1]

Figure 1 (b) shows resistance as a function of temperature for both cooling and heating cycles for a representative 1T-TaS$_2$ device with a channel thickness of $H = 60$ nm. The data reveal two hysteresis windows at the C-CDW – NC-CDW phase transition ($T = 200$ K) and NC-CDW – IC-CDW phase transition ($T = 355$ K), in line with previous reports[5]. The well-resolved hysteresis windows and their temperature range attest to the high-quality of the fabricated devices. In Figure 1 (c), we present resistance in the heating cycle in a wider temperature range (red curves) and normalized noise spectral density, $S_I/I^2$, at $f = 11$ Hz (blue curves) as the functions of temperature ($I$ is the current through the two-terminal device). The resistance is presented for three different 1T-TaS$_2$ devices (marked A, B and C) to illustrate reproducibility. The resistance and noise spectral density were measured at temperatures changing from low to high, with the heating rate of 2 K/min. The resistance and noise spectral density experience abrupt changes at the expected phase transition of 1T-TaS$_2$, *i.e.* C-CDW – NC-CDW near 200 K, NC-CDW – I-CDW near 350 K, and, finally, I-CDW – NP near 550 K. Note that the NP (normal phase) and M (metallic) symbols and terminology are used interchangeably in the text and figures.

The noise level decreases by two orders of magnitude, from $4 \times 10^{-9}$ Hz$^{-1}$ to $4 \times 10^{-11}$ Hz$^{-1}$, at the C-CDW – NC-CDW phase transition near 200 K, and it reveals a sharp, one order-of-magnitude peak at the NC-CDW – I-CDW phase transition near 350 K. The fact that we clearly observe the C-CDW – NC-CDW transition indicates that the selected 1T-TaS$_2$ thickness range is appropriate[23,38]. The abrupt changes in the noise spectral density happen at the same temperatures as the corresponding steps in the electrical resistance. Additional data confirming this conclusion are presented in the *Supplemental Information*. Our measurements in the low-bias regime prove that the noise spectroscopy can be used as an efficient tool for monitoring the phase transitions in a 1T-TaS$_2$ device[45].

Irrespective of the exact physical mechanism of the phase transitions in thin films of 1T-TaS$_2$, *i.e.* pure electric field effect, Joule self-heating or both, the electrical bias, $V_T$, required to induce the transitions should depend on the device geometry and design details. To induce the I-CDW – NP transition by the electric field, at reasonably low $V_T$, we fabricated devices with different geometries, channel lengths, and thicknesses. The strength of the applied field was varied from low to high (up to 35 kV/cm). The *I-V* characteristics and noise spectral density were measured as the functions of the electric field at different temperatures. Figure 2 shows the resistance and noise characteristics of a 1T-TaS$_2$ device with 60-nm-thick channel, measured at RT. The resistance as a function of the applied electric field, $E$, is presented in Figure 2 (a). It drops sharply at the field of 3.6 kV/cm, which corresponds to the NC-CDW – I-CDW phase transition. At $E = 6$ kV/cm, there is another relatively small step-like decrease in the resistance suggesting the I-CDW – NP transition. Although the change in the resistance is small, it is reproducible.

[Figure 2]

The normalized current noise spectral density, $S_I/I^2$, near the I-CDW – NP transition is shown in Figure 2 (b). The data are presented for different values of the electric field. Away from the phase-transition field, the noise spectra are of $1/f^\gamma$ type ($\gamma \approx 1$), indicating that there were no electromigration or dominant defect states, *i.e.* trap levels, with characteristic time constants. The noise increases with the trace of a Lorentzian peak at the transition bias, which corresponds to the onset of the I-CDW – NP transition. The evolution of the noise spectral density over the entire range of electric bias can be seen in Figure 2 (c), which shows $S_I/I^2$ ($f = 10$ Hz) as a function of electric field. There are three abrupt noise peaks. The peaks at $E = 3.6$ kV/cm and $E = 6$ kV/cm correspond to the NC-CDW – I-CDW and I-CDW – NP transitions, respectively. The noise level increases near or at the phase transition points owing to the lattice reconstruction. One should note that noise peak at the transition to the normal phase is much more clear than the slight step in the resistance at the same value of the electric field. The latter proves the usefulness of the noise spectroscopy for monitoring phase transitions. There is another peak in the noise spectral density at the applied field $E \approx 1$ kV/cm. We attribute this peak to the onset of CDW de-pinning, in line with our previous finding[45]. We also note that for some devices, the electric field to induce the phase transition, determined from the noise measurements, was somewhat lower than that from *I-V* characteristics. This difference can be attributed to the differences in sweep rates utilized for the *I-V* and noise measurements. This conclusion is in agreement with literature reports, which found a dependence of the NC-CDW – I-CDW transition on the voltage sweep rate[50,57]. Additional details supporting our conclusions, as well as data for different devices, are provided in the *Supplemental Materials*.

To investigate in detail the I-CDW – NP transition specifically, several 1T-TaS$_2$ devices were heated to 400 K, which is well above the NC-CDW – I-CDW phase transition temperature but substantially below the I-CDW – NP transition temperature of 550 K (see resistance data for device C in Figure 1). The tested devices had different channel lengths and widths in order to alter the electric bias, $V_B$, needed to induce the I-CDW – NP transition either through the field strength or Joule heating. The evolution of electric current through the device and the normalized noise spectra density in one of the devices are shown in Figures 3 (a), and 3 (b), respectively. The transition to metal phase is clearly seen as the step-like jump in the current, $I$, resistance, $R = V_B/I$, and as the peak in the normalized noise spectral density $S_I/I^2$. In this specific device, with larger dimensions, an electric field $E \approx 10.6$ kV/cm is required to induce the transition to the metallic state. Note a pronounced noise peak at this value of the electric field, which exactly corresponds to the field of the jump in current to a lower resistivity state (see inset to Figure 3 (c)). In other devices, the changes in the current were small, and the I-CDW – NP transition could be unambiguously determined only from the peak in the noise spectral density (see Figure 3 (c)).

[Figure 3]

A possibility of switching a CDW device to multiple resistive states by application of a small electric bias, *e.g.* below 3 V at RT, is important for information processing and memory applications[58,59]. To demonstrate switching among all three CDW phases and one normal phase, we set the 1T-TaS$_2$ device to the C-CDW state by cooling it below 200 K. In a device with properly selected dimensions, the application of a relatively small electric bias can induce all three transitions. Figure 4 (a) shows current as a function of the applied in-plane electric field at $T = 77$ K. The current shows three consecutive hysteresis windows at the electric fields of 25.0 kV/cm, 31.3 kV/cm, and 32.1 kV/cm, which correspond to the C-CDW – NC-CDW, NC-CDW – I-CDW, and I-CDW – NP transitions, respectively. Field-induced resistive switching is often dependent on the sweep rates because the CDW states require time to reorder, regardless of the mechanism. In Figure 4 (b), we present the current versus applied in-plane electric field at different voltage sweep rates, focusing on the $E > 30$ kV/cm range to better resolve the overlapping hysteresis region. One can see that the shapes of the hysteresis windows strongly depend on the scan rates.

[Figure 4]

The possibility of inducing all three transitions with applied electric bias can be further confirmed with the low-frequency noise measurements. Figure 5 (a) shows the noise spectral density, $S_I/I^2$, ($f = 10\ Hz$) as a function of the applied electric field for a different device. The measurements were conducted at $T = 77$ K to ensure that the C-CDW phase is the initial state of the device. The well-resolved peaks in the normalized noise spectral density at the field of 5 kV/cm, 8 kV/cm, and 12.5 kV/cm correspond the C-CDW – NC-CDW, NC-CDW – I-CDW, and I-CDW – NP transitions, respectively. One can notice that the noise spectral density, $S_I/I^2$, is a more accurate metric of all phase transitions than the resistivity, $R$ (see inset for comparison). It is interesting to note that the overall increase in the noise spectral density near the phase transition is often accompanied by the appearance of Lorentzian bulges in the noise frequency spectrum. In Figure 5 (b) and 5 (c), we present the noise spectral density as a function of frequency for several values of the electric field near the C-CDW – NC-CDW and NC-CDW – I-CDW transitions. The Lorentzian in Figure 5 (b) reveals itself as a $1/f^2$ shoulder because its central frequency is below the measurement range. The Lorentzian in Figure 5 (c) has a central frequency $f_o = 30\ Hz$.

[Figure 5]

The resistance changes near the I-CDW – NP transition can vary over a large range depending on the device structure and size. Figure 6 (a) shows the resistance in the "low-bias" regime near the I-CDW – NP transition induced by temperature. The resistance changes in two other devices induced by the applied in-plane electric field are presented in Figure 6 (b). Utilizing a large number of 1T-TaS₂ devices on Si/SiO₂ substrate with different channel geometry and sizes, we verified that switching among three CDW and one metallic phases can be achieved in a wide temperature range. Figure 6 (c) summarizes the results of the phase-transition switching with an applied electric filed. One can see that the spread of the required electric field can be as high $>$ 10 KV/cm. The electric field required for switching is determined by the device design and geometry, which affects the strength and distribution of electric field at a fixed bias voltage, and the dynamics of heat dissipation for a given input power. It is important to note that all the phase transitions considered here for 1T-TaS₂ can be induced by technologically feasible, small voltage biases. For most of the fabricated devices tested at RT, the I-CDW – NP transition was achieved at 2 V – 3 V bias, and the bias voltage required for the NC-CDW – I-CDW phase transition was substantially lower.

[Figure 6]

We now examine the role that Joule heating of the device channel plays in the phase transitions. First, we do a simple analytical model estimate of the temperature in a representative device with the lateral dimensions, $w \times L$, of 10 μm × 2 μm, and thickness $t_f = 50$ nm ($w$ and $L$ are the width and length of the channel, respectively). The device structure and *I-V* characteristics are shown in the *Supplementary Materials*. At room temperature, $T_{amb}$, the I-CDW–NP phase transition occurs at $V_d = 1.75$ V and $I_d = 17.5$ mA, which corresponds to the total Joule heating power of $P = V_d I_d = 30.6$ mW. The generated power is dissipated through the underlying SiO₂ layer with a thickness of $t_{ox} = 300$ nm and a thermal conductivity of $k_{ox} = 1.4$ Wm⁻¹K⁻¹. Although the in-plane thermal conductivity of the TaS₂ is approximately a factor of ×3 larger than that of the SiO₂ layer, owing to the small cross-section of the TaS₂ channel, $A_{f,\parallel} = t_f \times w$, the total in-plane thermal resistance of the TaS₂ layer, $R_{th,\parallel} = L/(k_f \times A_{f,\parallel})$ is about two orders of magnitude greater than the total cross-plane thermal resistance, $R_{th,\perp}$, of the underlying SiO₂/Si layers. For this reason, almost all dissipated heat diffuses through the SiO₂/Si substrate. The total cross-plane thermal resistance is a sum of three terms, $R_{th,\perp} = R_c + R_{SiO_2} + R_{Si}$, where $R_c$, $R_{SiO_2}$, and $R_{Si}$ are the thermal contact resistance at the TaS₂–SiO₂ interface, the thermal resistance of the SiO₂ layer, and the thermal resistance of the Si substrate, respectively. The thermal contact resistance at the interface and the thermal resistance of the SiO₂ layer are $R_c = \{G_{int} \times A_{f,\perp}\}^{-1}$ and $R_{SiO_2} = t_{ox}/(k_{ox} \times A_{f,\perp})$, where $A_{f,\perp} = w \times L$ is the device cross-plane surface area. The $G_{int}$ (interfacial thermal conductance) values reported for typical van der Waals materials on standard substrates vary in relatively small range and mostly depend on the quality of the interface[60–64], e.g. 50 MW/m² K for graphene, 33 MW/m² K for WTe₂, and 15 MW/m²K for MoS₂, respectively. An analytical estimate of the device average temperature rise across the SiO₂ layer is $\Delta T \approx P \times R_{SiO_2} \approx 328$ K, without taking into account the thermal contact resistance. This result suggests that the I-CDW–NP phase transition can take place due to the Joule heating, since the device temperature, $T = T_{amb} + \Delta T = 628$ K, exceeds the I-CDW–NP phase transition temperature of 550 K. Taking into account $R_c$ will add ~100–200 K to the temperature rise depending on the value for $G_{int}$ assumed for the device structure.

To further understand the phase transition mechanism and separate out the temperature effect from the field effect, we have modeled the heat diffusion in the experimentally tested sample, using the finite-element method implemented in COMSOL (see METHODS for details). The heat generation from Joule heating in the TaS$_2$ channel region is sensitive to the sample geometry, material properties, and ambient conditions. The effective electric field, $E$, and the current density, $J$, in the channel are taken to be same as in the experiment by tuning the sheet resistance of the TaS$_2$ active layer. The latter ensures that the Joule heating power, $P = V_d I_d$, corresponds to that of the experimental setup. We focus on the temperature distribution near two critical transition points: (i) just before IC-CDW – NP ($E \sim 8.45$ kV/cm), and (ii) just after IC-CDW – NP ($E \sim 9$ kV/cm). Since the current is localized between the inner contacts of the device structure, significant heating happens in the active region between two bias, *i.e.* source – drain, contacts. Our simulations show that just before the phase transition, the temperature of the hotspot can rise as high as $\sim 739$ K, well exceeding the C-CDW−NP transition temperature. After the phase transition, the temperature of the sample increases even more, up to $\sim 864$ K. The schematic of the simulated structure and temperature rise are provided in the Supplemental Materials.

Figure 7 (a)-(b) show the temperature profile along the cross-section of the device channel (see inset to Figure 7 (a)). The generated heat in the hotspots, indicated as red color regions at the middle of the channel, dissipates mainly through the SiO$_2$ layer and Si substrate. Owing to its low thermal conductivity, the SiO$_2$ layer acts as an effective thermal barrier, and prevents effective heat dissipation to the Si substrate. The latter results in substantial heating, which drives the phase transitions. The obtained numerical data for the specific device geometry is in agreement with the analytical estimate above. The radiation loss from the top surface becomes important at elevated temperature. However, it is not significant enough to change the conclusion that Joule heating is the reason behind the C-CDW−NP phase transition in TaS$_2$ devices of the considered geometry, on SiO$_2$/Si substrate.

[Figure 7]

In conclusion, we investigated switching among three charge density wave phases –

commensurate, nearly commensurate, incommensurate – and the normal metallic phase in thin films of 1T-TaS$_2$ induced by an in-plane electric field. The electric-field switching among all four phases has been achieved, which is a key prerequisite for nanoscale device applications. We have demonstrated that an application of small voltage bias to a device kept at room temperature can induce the transition from the nearly commensurate and incommensurate phases to the normal metallic phase. The low-frequency electronic noise spectroscopy was used as an effective characterization tool for monitoring the transitions, particularly the switching from the incommensurate charge-density-wave to the normal phase. The noise spectral density experiences sharp increases at the phase transition points, which correspond to the step-like changes in resistivity. The assignment of the phases has been confirmed by the low-field resistivity measurements over the temperature range from 77 K to 600 K. Analysis of the experimental data and simulations of heat dissipation suggests that self-heating plays a dominant role in the electric-field induced transitions. Furthermore, the results demonstrate the potential of the noise spectroscopy for investigating phase transitions in low-dimensional charge density wave materials.

**METHODS**

**Preparation of 1T-TaS$_2$ crystals by CVT:** Briefly, tantalum (20.4 mmol, Sigma-Aldrich 99.99% purity), sulfur (41.1 mmol, J.T. Baker >99.9% purity), and iodine (J.T. Baker 99.9% purity, ~88 mg for an ampule with a ~14.0 cm$^3$ volume) were added to a ~18×1 cm$^2$ cleaned, dried, fused quartz ampule. After evacuation, filling with Ar, and sealing, the ampule was heated in a two-zone tube furnace at 10 °C min$^{-1}$ to 975 – 875 °C (hot-cold gradient) for 1 week. The 1T-TaS$_2$ polymorph was isolated as golden, multi-millimeter-sized crystals by quenching the hot ampule in a water–ice–NaCl bath.

**Device fabrication:** We utilized the shadow mask method to directly deposit TLM structures onto pre-selected 1T-TaS$_2$ thin films obtained by mechanical exfoliation. The exfoliated layers of 50 nm – 60 nm were relatively stable in the air, which allowed us to use the shadow mask method[65]. The method allowed us to avoid the damage from chemical contamination, typically associated with conventional lithographic lift-off processes. It also drastically reduced the total air exposure time and the fabrication process. It takes less than an hour to fabricate a device by the shadow mask method, compared to 2-3 days required for the standard lithography process. The designed shadow masks of TLM configuration were fabricated using the double-side polished Si wafers with 3 μm thermally grown SiO$_2$ on both sides (Ultrasil Corp.; 500 μm thickness; P-type; <100>). The shadow mask fabrication process began with evaporation of 200 nm thick chromium (Cr) film on the front side of the wafer, followed by a combination of electron beam lithography (EBL) and Cr etchant (1020A) to create TLM pattern. This was followed by the fluorine-based reactive ion etching (RIE) to transfer the pattern to the underlying SiO$_2$. Finally, the pattern was transferred into the underlying Si substrate using the deep reactive ion etching (DRIE) (Oxford Cobra). Since the 500 μm thick Si wafer is too bulky to etch through from the top, a large window was opened at the back side of the wafer by photolithography to remove the bulk Si until a thin Si membrane is left. Finally, The DRIE etch step was timed to break through the thin silicon membrane and release the stencil from the top. The shadow masks were directly used to fabricate plenty of devices by aligning them with pre-selected flakes on the device substrate using an optical microscopy, clamping the aligned mask and device substrate together, and placing the clamped assembly in an electron beam evaporator (EBE) for contact deposition (10 nm Ti and 100 nm Au) through the mask openings. The completed devices were then transferred to another vacuum chamber for electrical characterization. The thickness and width of 1T-TaS$_2$ flakes were determined using atomic force microscopy (AFM) and scanning electron microscopy (SEM) inspection.

**Electronic noise measurements:** The noise spectra were measured with a dynamic signal analyzer (Stanford Research 785). To minimize the 60 Hz noise and its harmonics, we used a battery biasing circuit to apply voltage bias to the devices. The devices were connected to Lakeshore cryogenic probe station TTPX. All *I-V* characteristics were measured in the cryogenic probe station (Lakeshore TTPX) with a semiconductor analyzer (Agilent B1500). The noise measurements were conducted in the two-terminal device configuration but with different channel length to determine the contribution of contact noise. Since the contact resistance of the devices are very low, the extracted contact noise was lower than thermal noise and played has no significant contribution to the noise due to the channel. The dynamic signal analyzer measures the absolute voltage noise spectral density $S_V$ of a parallel resistance network of a load resistor, $R_L$ (3.6 kΩ), and device under test (DUT), $R_D$. The normalized current noise spectral density, $S_I/I^2$, is calculated from the Electric Switching of the Charge-Density-Wave and Normal Metallic Phases in 1T-TaS$_2$ Thin-Film Devices – UC Riverside 2019equation: $S_I/I^2 = S_V \times [(R_L + R_D)/(R_L \times R_D)]^2 / (I^2 \times G^2)$, where $G$ is the amplification of the low-noise amplifier (1000).

**Finite element simulations:** To model the heat distribution in the TaS$_2$ channel, we used the finite element model, implemented in COMSOL Multiphysics package[66]. The simulated device geometry corresponds to an experimental sample, which has a 50 nm thick TaS$_2$ sample on top of a 4 µm thick Silicon substrate. Four 2.5 µm × 10 µm Au electrodes are placed on top of TaS$_2$, with inter-contact separations of 2 µm. Bias is applied between the two middle Au contacts. In the experiment, the C-CDW−NP transition happens for $V = 1.69$ V (19.25 mA) and 1.80 V (24.58 mA). Same electrical conditions are maintained in the simulation geometry. In case of thermal conduction, heat dissipation can be described by standard heat equation,

$$\rho C_p \frac{dT}{dt} = \nabla(\kappa \nabla T) + q_e, \qquad (2)$$

where, $\rho$ is the mass density, $C_p$ is the heat capacity, $T$ is temperature, and $\kappa$ is thermal conductivity. The left side of Eq. (2) will disappear under steady state condition. On the right side of Eq. (2), Joule heating is captured by the generation term, $q_e = J \cdot E = \sigma |\nabla V|^2$, where $J$ is the current density, $E$ is the electric field, $\sigma$ is the electrical conductivity, and $V$ is the applied bias. The generated heat will appear as an outward heat flux, giving rise to a thermal gradient, $\nabla T$. The electrical conductivity of the TaS$_2$ channel ($\sigma$) is tuned to match the current and power density of the experimental sample. The thermal conductivity of TaS$_2$ ($\kappa$) is around $4 \sim 10$ Wm$^{-1}$K$^{-1}$[Ref. 67]. For SiO$_2$ layer, the thermal conductivity is 1.4 Wm$^{-1}$K$^{-1}$[Ref. 68]. The metal contacts, as well as the bottom face of the Si substrate, are assumed to be always at room temperature (300 K). Radiating heat boundary conditions are applied on the exposed TaS$_2$ and SiO$_2$ surfaces as, $\hat{n} \cdot \kappa \nabla T = \varepsilon \sigma (T_{env}^4 - T^4)$, where $T_{env}$ is the ambient temperature (300 K), $\hat{n}$ is the surface normal and $\varepsilon$ is the emissivity (0.8). All other surfaces are assumed to be thermally insulating ($\hat{n} \cdot \kappa \nabla T = 0$).

**Acknowledgements**

The work at UCR and UGA was supported, in part, by the National Science Foundation (NSF) through the Emerging Frontiers of Research Initiative (EFRI) 2-DARE project: Novel Switching Phenomena in Atomic MX$_2$ Heterostructures for Multifunctional Applications (NSF EFRI-14 | P a g e

1433395). A.A.B. also acknowledges the UC - National Laboratory Collaborative Research and Training Program - University of California Research Initiatives LFR-17-477237. Nanofabrication was performed in the Center for Nanoscale Science and Engineering (CNSE) Nanofabrication Facility at UCR. S. R. acknowledges partial support from the Center for Terahertz Research and Applications co-financed with the European Regional Development Fund.

**Contributions**

A.A.B. conceived the idea, coordinated the project, contributed to experimental data analysis, and led the manuscript preparation; A.G. fabricated the devices, conducted current-voltage and noise measurements, contributed to data analysis; S.R. supervised noise measurements and contributed to data analysis; B.D. conducted simulations and contributed to data analysis; F.K. contributed to analysis of experimental and simulation data and performed analytical derivations; A.N. tested high-temperature current-voltage characteristics; R.K.L. supervised numerical simulations and contributed to data analysis; M.B supervised high-temperature measurements. M.B. synthesized 1T-TaS$_2$ crystals and conducted materials characterization; T.T.S. supervised material synthesis and contributed to materials characterization. All authors contributed to writing the manuscript.

Electric Switching of the Charge-Density-Wave and Normal Metallic Phases in 1T-TaS$_2$ Thin-Film Devices – UC Riverside 2019Structure in 1 T-TaS$_2$. *Science* **1992**, *257* (5068), 362–364.

(11) Wu, X. L.; Lieber, C. M. Hexagonal Domain-Like Charge Density Wave Phase of TaS$_2$ Determined by Scanning Tunneling Microscopy. *Science* **1989**, *243* (4899), 1703–1705.

(12) Carpinelli, J. M.; Weitering, H. H.; Plummer, E. W.; Stumpf, R. Direct Observation of a Surface Charge Density Wave. *Nature* **1996**, *381*, 398.

(13) Vogelgesang, S.; Storeck, G.; Horstmann, J. G.; Diekmann, T.; Sivis, M.; Schramm, S.; Rossnagel, K.; Schäfer, S.; Ropers, C. Phase Ordering of Charge Density Waves Traced by Ultrafast Low-Energy Electron Diffraction. *Nature Physics* **2017**, *14*, 184.

(14) Grüner, G. The Dynamics of Charge-Density Waves. *Reviews of Modern Physics* **1988**, *60* (4), 1129–1181.

(15) Fazekas, P.; Tosatti, E. Electrical, Structural and Magnetic Properties of Pure and Doped 1T-TaS$_2$. *Philosophical Magazine B* **1979**, *39* (3), 229–244.

(16) Kim, J.-J.; Yamaguchi, W.; Hasegawa, T.; Kitazawa, K. Observation of Mott Localization Gap Using Low Temperature Scanning Tunneling Spectroscopy in Commensurate 1T-TaS$_2$. *Physical Review Letters* **1994**, *73* (15), 2103–2106.

(17) Zwick, F.; Berger, H.; Vobornik, I.; Margaritondo, G.; Forró, L.; Beeli, C.; Onellion, M.; Panaccione, G.; Taleb-Ibrahimi, A.; Grioni, M. Spectral Consequences of Broken Phase Coherence in 1T-TaS$_2$. *Physical Review Letters* **1998**, *81* (5), 1058–1061.

(18) Fazekas, P.; Tosatti, E. Charge Carrier Localization in Pure and Doped 1T-TaS$_2$. *Physica B+C* **1980**, *99* (1), 183–187.

(19) Mott, N. F. Metal-Insulator Transition. *Reviews of Modern Physics* **1968**, *40* (4), 677–683.

(20) Wilson, J. A. Questions Concerning the Form Taken by the Charge-Density Wave and the Accompanying Periodic-Structural Distortions in 2H-TaSe$_2$ and Closely Related Materials. *Physical Review B* **1978**, *17* (10), 3880–3898.

(21) Spijkerman, A.; de Boer, J. L.; Meetsma, A.; Wiegers, G. A.; van Smaalen, S. X-Ray Crystal-Structure Refinement of the Nearly Commensurate Phase of 1T-TaS$_2$ in (3+2)-Dimensional Superspace. *Physical Review B* **1997**, *56* (21), 13757–13767.17 | P a g e

**CAPTIONS**

**Figure 1:** (a) Optical image of a representative device (left panel) and a schematic of the device layered structure (right panel). The red and blue dashed lines show the contours of the 1T-TaS$_2$ channel and a metal contact, respectively. The scale bar is 2 µm. (b) Resistance as function of temperature for cooling (blue curve) and heating (red curve) cycles conducted at the rate of 2 K per minute. The resistance reveals two characteristic hysteresis windows at the commensurate – nearly commensurate CDW transition at $T \approx 200$ K and at the nearly commensurate – incommensurate CDW transition at $T \approx 355$ K. (c) Resistance (red curves) and normalized noise spectral density (blue curve), as function of temperature, measured in the heating cycle. The resistance is presented for three different 1T-TaS$_2$ devices (marked A, B, C). The noise spectral density, measured at $f = 10$ Hz, is for device A. The device C was tested for higher temperatures to resolve the transition from incommensurate CDW to the normal metallic phase ($T \approx 550$ K), marked as IC-M in the legend. All three devices show an abrupt resistance change at $T \approx 355$ K due to the transition from nearly commensurate to incommensurate CDW phase. Two devices A and B, tested at low temperature, show an abrupt resistance change at $T \approx 200$ K due to the transition from the commensurate to nearly commensurate CDW phase.

**Figure 2:** (a) Resistance as a function of the applied electric field for 1T-TaS$_2$ device measured at room temperature. The resistance drops sharply at $E = 3.6$ kV/cm owing to the nearly commensurate to incommensurate CDW phase transition, marked as NC-IC in the legend. The slight step in resistance at $E = 6.0$ kV/cm corresponds to the transition from incommensurate CDW phase to the normal metallic phase, marked as IC-M in the legend. (b) Normalized current noise spectral density as the function of frequency for several values of the electric field, which include the point of transition from the incommensurate CDW to the normal metallic phase. The noise level is the lowest in the metallic phase. It increases strongly at the electric field value, close to the one required for inducing the transition to the metallic state. (c) Normalized current noise spectral density, measured at $f = 10$ Hz, as the function of the electric field. One can clearly see two peaks, which correspond to the NC-IC and IC-M phase transitions at $E = 3.6$ kV/cm and $E = 6.0$ kV/cm, respectively. An additional peak at $E \approx 1.0$ kV/cm is attributed to the de-pinning of the charge density wave.

**Figure 3:** (a) Current through the device channel as a function of the applied electric field. The data were measured for the device intentionally heated to $T = 400$ K to ensure that 1T-TaS$_2$ is in the incommensurate CDW phase. One can see a sharp increase in the current at the transition from the incommensurate CDW to normal metallic phase. (b) The normalized current noise spectral density, measured at $f = 10$ Hz, as a function of the electric field for the same device and temperature as in (a). The noise spectral density reveals a well-resolved peak at the transition point, marked as IC-M in the legend. (c) The normalized current noise spectral density, measured at $f = 10$ Hz, as a function of the electric field for a different device with larger channel dimensions. The well-resolved peak in the noise spectral density, again, clearly identifies the incommensurate CDW – metallic phase transition point, marked as IC-M in the legend. The inset shows the current for this device with the distinguishable change at the transition point to the metallic phase. The data attests to the potential of the low-frequency noise spectroscopy for monitoring phase transitions in CDW materials.

**Figure 4:** (a) Current through the device channel as a function of the applied bias electric field. The data were measured for the device intentionally cooled to $T = 77$ K to ensure that 1T-TaS$_2$ is in the commensurate CDW phase. The current dependence on the field shows three consecutive hysteresis windows at $E = 25.0$ kV/cm, $E = 31.3$ kV/cm and $E = 32.1$ kV/cm, which correspond to the commensurate CDW to the nearly commensurate CDW, nearly commensurate CDW to incommensurate CDW, and incommensurate CDW to metallic phases of 1T-TaS$_2$, respectively. (b) Current though the same device measured in high-field region at different bias voltage sweep rates. The slower sweep rates allow for better resolution of the phase transitions. Note the sweep-rate dependence of the field, required to induce the transitions.

**Figure 5:** (a) Normalized current noise spectral density measured at $f = 10$ Hz as a function of electric field applied to the device kept at 77 K. The noise spectral density reveals three well defined peaks at the electric field of 5 kV/cm, 8 kV/cm and 12.5 kV/cm, which correspond to the transitions from the commensurate to nearly commensurate, from the nearly commensurate to incommensurate, and from the incommensurate to normal metallic phases, respectively. The inset shows the channel resistance as a function of the electric field for the same device. The transition points are indicated with the arrows. (b) The noise spectral density as a function of frequency for several values of the electric field near the commensurate to nearly commensurate phase transition. The Lorentzian reveals itself as the $1/f^2$ shoulder because its central frequency is below the measurement range. (c) The same as in (b) but for the electric field, which correspond to the transition from the nearly commensurate to incommensurate phase.

**Figure 6:** (a) Device channel resistance as a function of temperature near the transition from the incommensurate charge-density-wave phase to normal metallic phase. (b) Resistance as a function of applied electric field near the transition from the incommensurate charge-density-wave phase to normal metallic phase. The data are shown for two other devices with smaller geometries. (c) Summary of electric field induced phase transitions at different temperatures for 1T-TaS$_2$ devices. The variation in the electric field required to include the phase transitions is due to different device geometries, thickness of the layers in the device structures, and other variations in the device designs.

**Figure 7:** (a) Simulated temperature distribution profile along the cross-section of the device channel for the voltage bias just before the C-CDW-NP transition. The hot spot temperature exceeds 550 K, which is above the C-CDW-NP transition. (b) The same as in (a) but for the voltage bias, which corresponds to the regime above the C-CDW-NP phase transition. The temperature increases owing to the lower resistance of 1T-TaS$_2$ in the metallic NP, which results in stronger Joule heating.

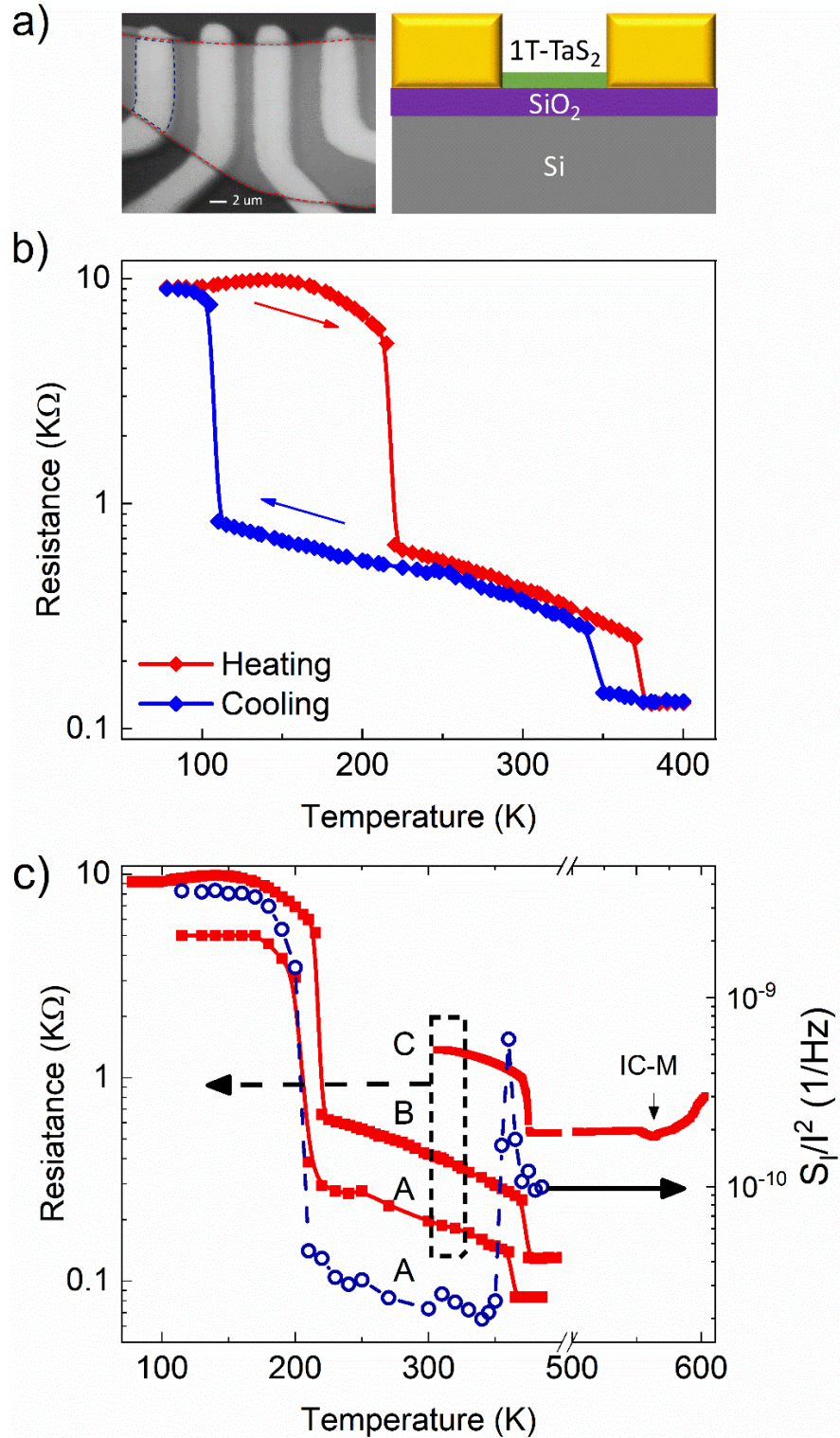

**Figure 1**

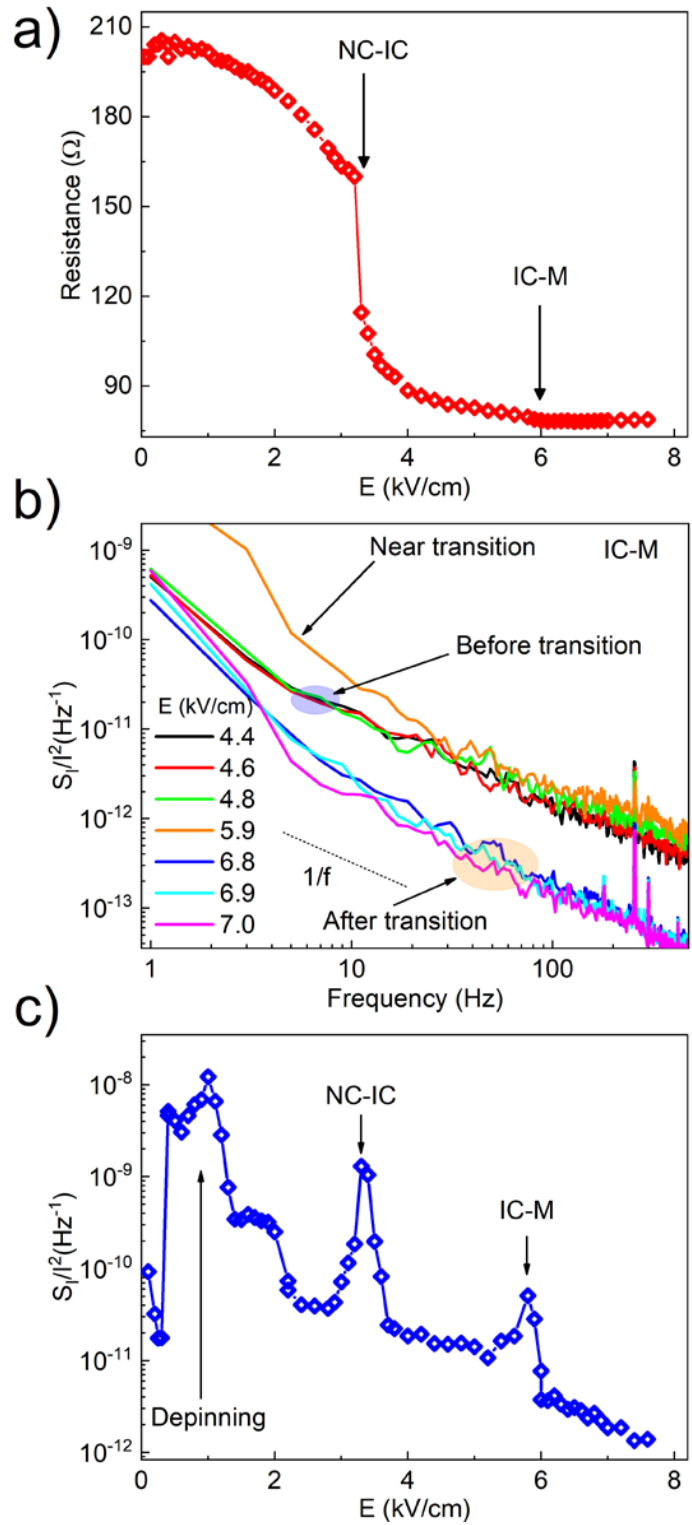

**Figure 2**

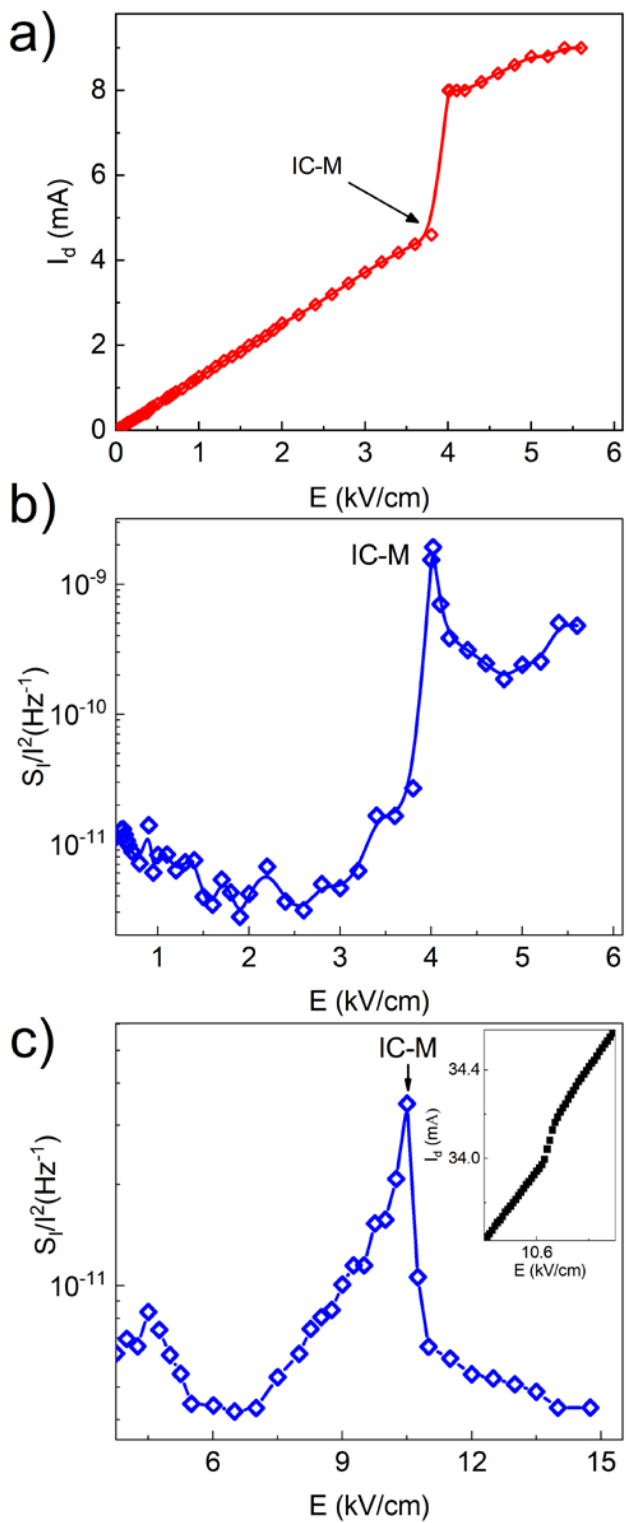

**Figure 3**

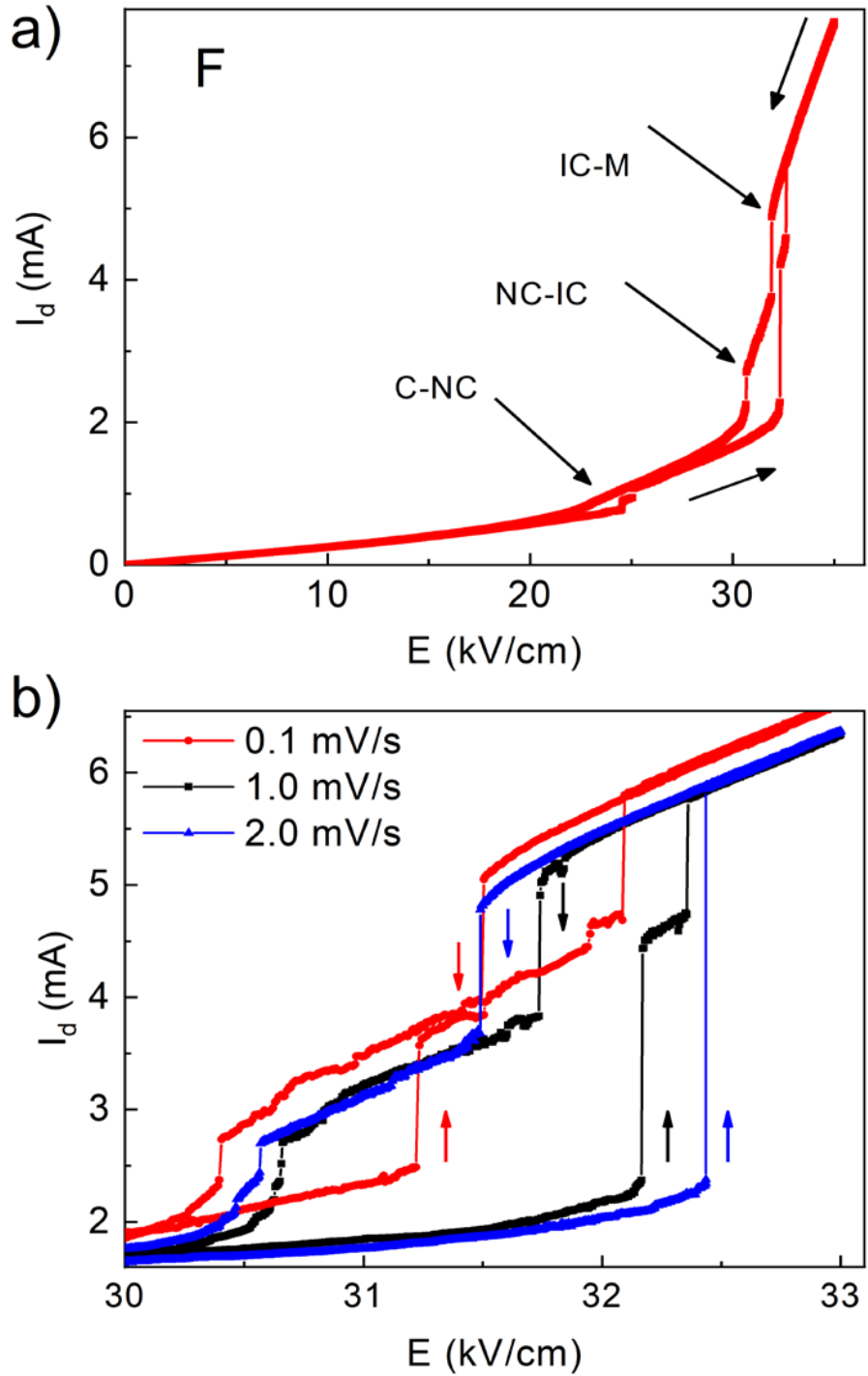

**Figure 4**

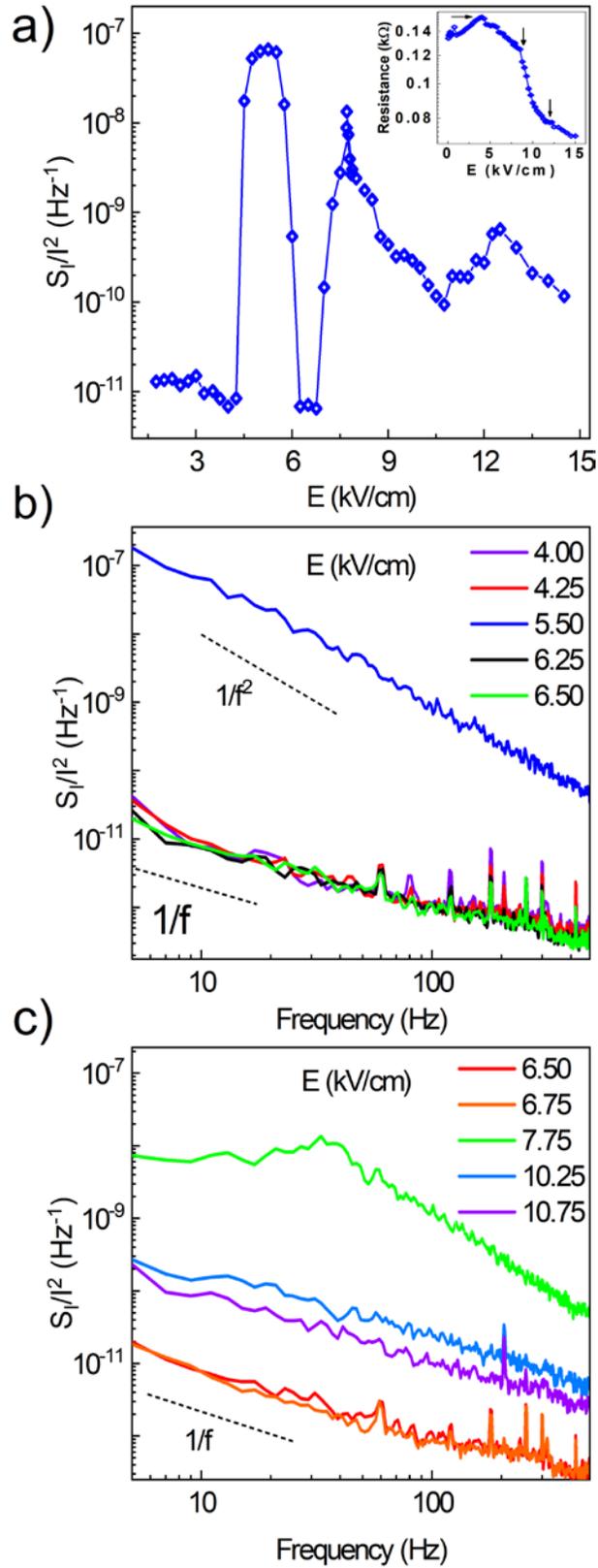

**Figure 5**

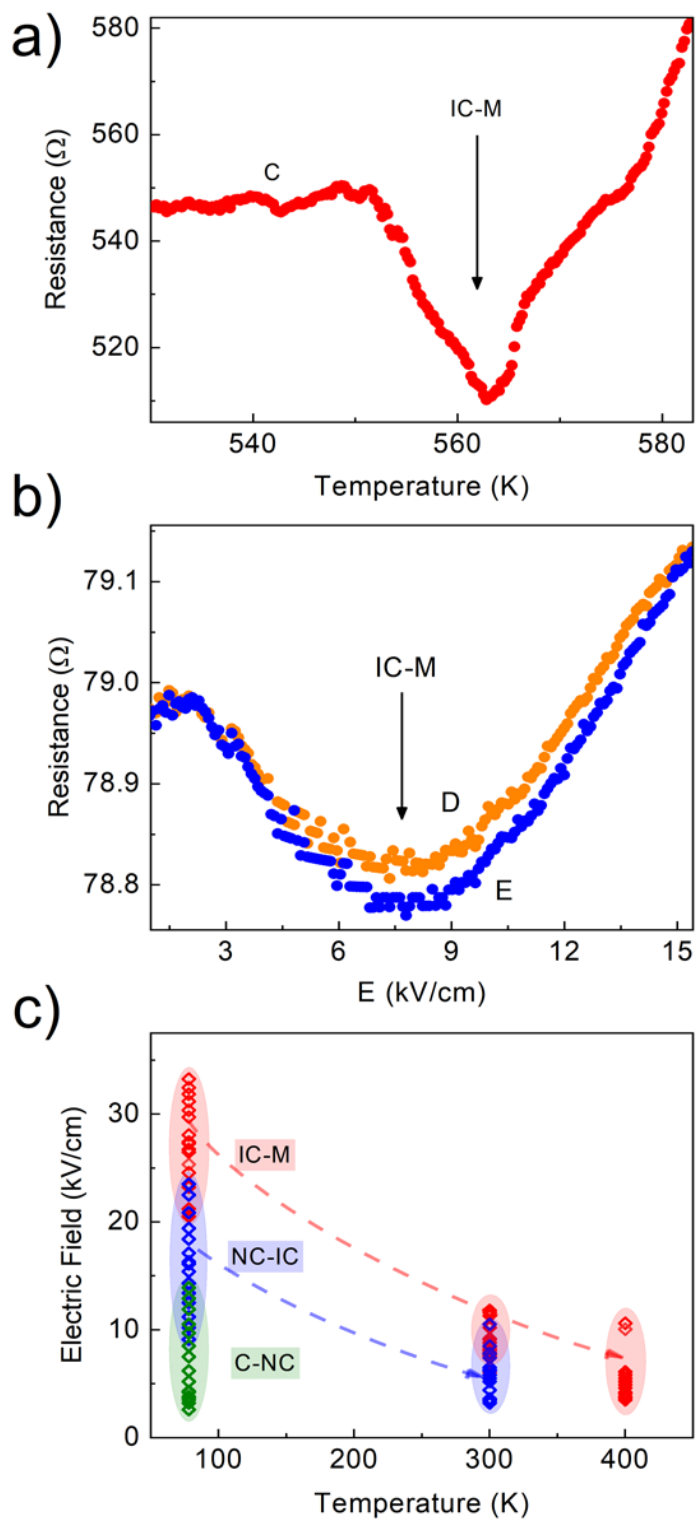

**Figure 6**

Electric Switching of the Charge-Density-Wave and Normal Metallic Phases in 1T-TaS$_2$ Thin-Film Devices – UC Riverside 2019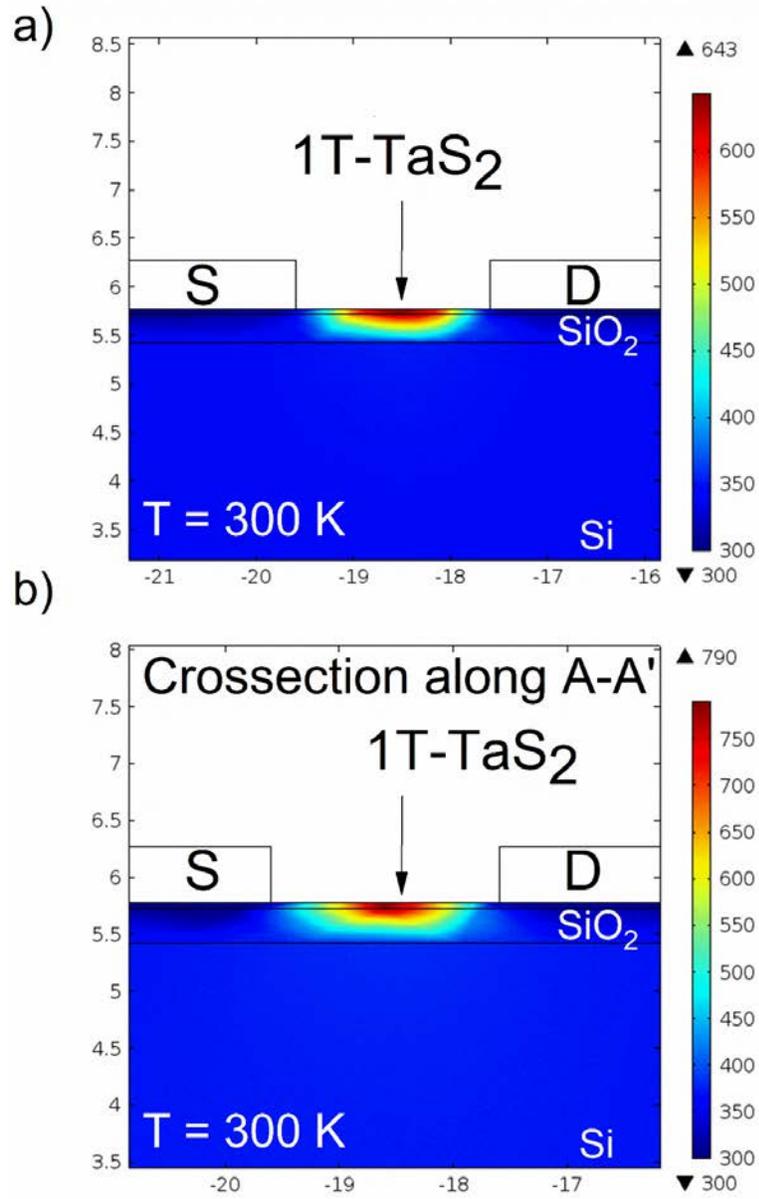

**Figure 7**

32 | P a g e